\documentclass[11pt]{article}
\usepackage{amsmath,amssymb,color,graphicx,subfigure,float}

\usepackage{amsfonts}

\newcommand{\hoch}[1]{$\, ^{#1}$}

\makeatletter
\@addtoreset{equation}{section}
\makeatother

\newcommand{\be}{\begin{equation}}
\newcommand{\ee}{\end{equation}}
\newcommand{\bea}{\setlength\arraycolsep{2pt} \begin{eqnarray}}
\newcommand{\eea}{\end{eqnarray}}
\newcommand{\nn}{\nonumber}
\newcommand{\ben}{\begin{equation}}
\newcommand{\een}{\end{equation}}

\def\ft#1#2{{\textstyle{\frac{\scriptstyle #1}{\scriptstyle #2} } }}
\def\fft#1#2{{\frac{#1}{#2}}}
\def\CP{{{\mathbb C}{\mathbb P}}}
\def\0{{\sst{(0)}}}
\def\1{{\sst{(1)}}}\def\2{{\sst{(2)}}}
\def\3{{\sst{(3)}}}
\def\4{{\sst{(4)}}}
\def\5{{\sst{(5)}}}
\def\6{{\sst{(6)}}}
\def\7{{\sst{(7)}}}
\def\8{{\sst{(8)}}}
\def\sst#1{{\scriptscriptstyle #1}}

\def\del{{\partial}}
\def\im{{{\rm i}}}

\def\cA{{{\cal A}}}
\def\cF{{{\cal F}}}
\def\cM{{{\cal M}}}

\def\cV{{{\cal V}}}

\def\p{\partial}
\def\vep{{\varepsilon}}
\def\tvep{{\tilde\varepsilon}}

\begin{document}

\begin{flushright}
\hfill{ \
DAMTP-2013-5\ \ \ \ \ MIFPA-13-02\ \ \ \ }
\end{flushright}

\vspace{25pt}
\begin{center}
{\Large {\bf Ergoregions in Magnetised Black Hole Spacetimes}}

\vspace{35pt}

{\Large G.W. Gibbons\hoch{1}, A.H. Mujtaba\hoch{2} and C.N. Pope\hoch{1,2}}

\vspace{15pt}

\hoch{1}{\it DAMTP, Centre for Mathematical Sciences,
 Cambridge University,\\  Wilberforce Road, Cambridge CB3 OWA, UK}

\vspace{10pt}

\hoch{2} {\it George P. \& Cynthia Woods Mitchell  Institute
for Fundamental Physics and Astronomy,\\
Texas A\&M University, College Station, TX 77843, USA}

\vspace{20pt}

\underline{ABSTRACT}
\end{center}

  The spacetimes obtained by Ernst's procedure for appending an 
external magnetic field $B$ to a seed Kerr-Newman black hole are commonly
believed to be asymptotic to the static Melvin metric.  We show that this
is not in general true. Unless the electric charge of the black hole
satisfies $Q= jB(1+\ft14 j^2 B^4)$, where $j$ is the angular momentum of
the original seed solution, an ergoregion extends all the way from the black
hole horizon to infinity.  We find that if the condition on the electric 
charge is satisfied then the metric is asymptotic to the static Melvin metric,
and the electromagnetic field carries not only magnetic, but also electric,
flux along the axis.
We give a self-contained account of the
solution-generating procedure, including explicit formulae for
the metric and the vector potential.  In the case when 
$Q= jB(1+\ft14 j^2 B^4)$, we show that there is an arbitrariness in the
choice of asymptotically timelike Killing field $K_\Omega= \del/\del t+
\Omega\, \del/\del \phi$, because there is no canonical choice of $\Omega$.
For one choice, $\Omega=\Omega_s$, the metric is asymptotically static,
and there is an ergoregion confined to the neighbourhood of the 
horizon.  On the other hand, by choosing $\Omega=\Omega_H$, so that 
$K_{\Omega_H}$ is co-rotating with the horizon, then for sufficiently
large $B$ numerical studies indicate 
there is no ergoregion at all.  For smaller values, in a range
$B_-<B<B_+$, there is a toroidal ergoregion outside and disjoint from the
horizon. If $B\le B_-$ this ergoregion expands all the way to infinity
in a cylindrical region near to the rotation axis.
For black holes whose size is
small compared to the Melvin radius $2/B$, and neglecting back-reaction of
the electromagnetic field, we recover Wald's result that it
is energetically favourable for the hole to acquire a charge $2jB$.

\vspace{15pt}

\pagebreak
\voffset=-40pt
\setcounter{page}{1}

\tableofcontents

\addtocontents{toc}{\protect\setcounter{tocdepth}{2}}


\section{Introduction}

  Understanding  the energetics of  astrophysical black holes involves
understanding their  interactions with charged particles and
with external magnetic fields.  This has been the subject of many studies, 
going back to the early work
of Wald \cite{Wald}, King, Kundt and Lasota \cite{KKL}, 
Blandford and Znajek \cite{BlandfordZnajek}, 
etc. (see \cite{Bicak1,Bicak2,Komissarov} for recent reviews). 
In particular, when the black hole
is rotating, the resulting ``dragging\rq \rq of the magnetic field
induces  electric fields which may have dramatic effects on
charged particles,  to the extent that it becomes energetically favourable
for an initially neutral black hole of mass $m$ and angular momentum
$j$ to acquire  a charge \cite{Wald}, and for currents to flow
\cite{BlandfordZnajek}.    

For all practical astrophysical  purposes, the gravitational 
back-reaction of the magnetic field may be neglected, and the
electromagnetic field may be treated as a \lq\lq test\rq \rq
field on the  unperturbed, asymptotically flat,
 electrically neutral  Kerr solution, with mass parameter
$m$ and angular momentum factor $j=am$. If the electromagnetic field
is assumed to be stationary, one may then use an old result
of Papapetrou \cite{Papapetrou} to show that the vector potential takes 
the form \cite{Wald}
\ben
A = \Big( \frac{jB}{m} -\frac{Q}{2m} \Big)
 K_\flat +  \ft12 B {\mathfrak m}_\flat \,,
\een    
where $K_\flat= K^\mu g_{\mu \nu} dx^\nu$ and ${\mathfrak m}_\flat =
 {\mathfrak m}^\mu g_{\mu \nu} dx^\nu$. 
Here, $ K^\mu \p_\mu= \frac{\p}{\p t}$ and 
$ {\mathfrak m}^\mu \p_\mu= \frac{\p}{\p \phi}$ are the time-translation and 
rotational Killing fields respectively, the constant $B$ is the 
strength of the asymptotic magnetic field, and the constant
$Q$ is the charge inside the horizon.    The electrostatic 
potential difference $\Phi_H$, or ``injection energy''
  between  the black hole horizon and infinity
is given by
\ben
\Phi_H =  \frac{Q-2jB}{2m} \qquad 
\een
and will vanish if the hole acquires the {\it Wald charge} 
\be
Q=2jB\,.\label{wald}
\ee
The mechanism for the  current flow required to lower the energy
might be conduction 
through the ambient plasma, or a breakdown of the vacuum through 
pair production. A discussion of  pair production from the point of view
of black hole thermodynamics and quantum field theory 
was  given long ago  \cite{Gibbons}; 
however, back-reaction was not then taken into 
account.  More recently, Wald's original argument 
has been criticised by Li \cite{Li}, who proposed a different value
for  the charge which minimizes the electromagnetic energy.  

Since those early investigations, despite the availability
of apparently appropriate  exact solutions of the Einstein-Maxwell equations
taking back reaction into account \cite{Ernst3,ErnstWild}, and an analysis
of their properties \cite{Galtsov1,Aliev1,Dokuchaev,Aliev2,Aliev3,Aliev4},  
no full treatment of black hole  thermodynamics in the rotating
case has yet been given. An appealing analogy to a Faraday disc, adopted by 
Blandford and Znajek, based on  ideas of Damour \cite{Damour}, views
the horizon as an electrical 
conductor with surface conductivity of  $4\pi$ Ohms.  This was 
elaborated upon by Thorne et al., 
under the rubric of the \lq\lq Membrane Paradigm\rq \rq \cite{Thorne}.
It suggests that a full treatment, taking into account back-reaction
and in particular the torque exerted by the rotating black hole
on the source of the magnetic field, might be extremely rich.
This expectation gains support from the
striking fact that  in the only case that has to date been studied exactly,
namely that of a Schwarzschild black hole immersed  in a background Melvin
solution \cite{Melvin} (possibly with a dilaton in addition),
it was found that the black hole thermodynamics
was unaffected by the presence of the magnetic field, with both
the entropy and temperature being unchanged \cite{Radu}. There
is a clear implication of this result that the  microscopic
degrees of freedom of the hole that are responsible for the entropy
are unaffected by the external magnetic field.  The 
obvious question arises as to whether this remains true in the rotating case. 

The reason for the absence of a full treatment is the
complexity of the exact solutions that appear to be appropriate in this case, 
all of which  have been obtained by means of Harrison type  solution-generating 
techniques \cite{Harrison} starting from an initial Kerr-Newman metric.
The default assumption in the literature has been that this
will produce a background at infinity that is ``asymptotically Melvin.'' 
If this were the case, 
it should then be a straightforward task to read off 
the total mass and angular momentum, calculate
the electrostatic potentials, and hence get a handle on the
generalised first law, possibly using Komar identities.
So far, the complexity of the solutions, even at infinity,
and possibly the absence  of a direct astrophysical motivation,     
has prevented this programme being carried out.
In this paper we shall show that there is a more serious
obstruction:  the relevant metrics turn out in general not to be
asymptotic to the static Melvin metric. In fact, unless the
charge parameter $q$ of the magnetised Kerr-Newman solution is
chosen to be  $q=-a m B$, where $a$ is the rotation parameter, $m$ the mass
parameter and $B$ the external magnetic field, they contain an ergoregion
that extends out to  infinity, close to, but not containing, the
rotation axis, with timelike boundary. 
In other words, unless $q=-am B$, {\sl the dragging of inertial 
frames is so strong that
even at infinity there is no Killing vector field which
is  everywhere timelike outside a compact set containing the black hole
Killing horizon}.   If $q=-amB$, we find that the metric is asymptotic to
the Melvin metric, but the electromagnetic field contains both electric
and magnetic components that are asymptotically constant on the axis. The
$q=-amB$ 
solution is in fact asymptotic to a duality rotation of the Melvin magnetic
universe.

  The criterion $q=-amB$ that the ergoregion does not extend to infinity may be
re-expressed in terms of the total electric charge $Q$ and the angular
momentum $j=am$ of the original seed solution:
\be
Q = jB (1+\ft14 j^2 B^4)\,.\label{spec}
\ee
Note that the quantity $j$ should be distinguished from any measure of the
total angular momentum of the magnetised spacetime.  For asymptotically
flat spacetimes, the total angular momentum $J$ may be expressed as a
Komar integral
\be
J = \fft1{16\pi} \int {*d}{\frak m}_\flat\,,\label{komar}
\ee
taken over a large sphere at spatial infinity.  For a vacuum spacetime,
such integrals do not depend on the choice of surface on which they are
evaluated, provided the surface is homologous to the sphere at infinity.
In the presence of an electromagnetic field, the Komar integral may be
surface dependent.  For example, in the case of a Kerr-Newman black hole,
the total angular momentum $j$ is given by a Komar integral over a surface
at infinity, and this differs from the integral over the horizon because of
the angular momentum carried by the electromagnetic field outside the 
horizon.  In the case of the magnetised spacetimes, difficulties arise
in trying to evaluate (\ref{komar}) because of the
asymptotic structure.

   One may check that the existence of an ergoregion that extends to infinity,
in the case that (\ref{spec}) is not satisfied, is independent of the
choice of timelike Killing vector.  More specifically, if one replaces
$K=\del/\del t$ by $K_\Omega=
\del/\del t + \Omega \del/\del\phi$, where $\Omega$
is a constant, then $K_\Omega$ 
will still be spacelike at infinity, close to
the rotation axis.

  If the charge does take the special value given by $q=-amB$ (or,
equivalently, (\ref{spec})), we find that there exists a range of choices for
$\Omega$, of the form $\Omega_- < \Omega <\Omega_+$, for which the Killing
vector $K_\Omega$ is timelike everywhere at large distances.  Within this
range lies an angular velocity $\Omega_s$ for which the magnetised black
hole metric is asymptotically static.  In this frame, $K_\Omega$ 
becomes spacelike in a compact neighbourhood of the horizon, signaling
the occurrence of an ergoregion that is similar to the one outside a
standard Kerr or Kerr-Newman black hole.  For another choice of $\Omega$
within the range, namely $\Omega=\Omega_H$, the angular velocity of the
horizon, the Killing vector $K_\Omega$ is null on the horizon and,
for sufficiently large $B$ (greater than a certain critical value $B_+$), 
numerical studies indicate it is timelike
everywhere outside the horizon.  Thus in this frame, there is no ergoregion
at all when $B>B_+$.  If $B$ lies in the range $B_-<B<B_+$, where
$B_-$ is another computable value of the magnetic field, there is an
ergoregion of toroidal topology, outside the horizon and disjoint from it,
lying in the equatorial plane.  As $B$ approaches $B_-$ from above, the
toroidal ergoregion extends upwards and downwards further and further,
eventually reaching infinity if $B\le B_-$.  

   The possibility of making different choices for the asymptotically
timelike Killing vector that generates ``time translations'' is something
that does not arise in asymptotically flat stationary spacetimes,
where there is a unique asymptotically timelike Killing vector.  We 
include in this paper a detailed discussion of this phenomenon in 
asymptotically Melvin spacetimes, and make a comparison with the
somewhat analogous situation that arises for stationary black holes in
asymptotically AdS spacetimes. 

   The organisation of this paper is as follows.  As a preliminary step, 
before considering the full magnetised Kerr-Newman solution, in section 2 we
examine the much simpler case of the magnetised Reissner-Nordstr\"om 
black hole. This example is useful because it illustrates, in a simpler
setting, the same problem that arises for a general magnetised Kerr-Newman
black hole, namely, that there is no choice of Killing vector that is
asymptotically timelike in all directions at infinity.  Specifically,
we find that near to the $z$ axis any Killing vector of the form 
$K=\del/\del t+ \Omega\, \del/\del\phi$ becomes spacelike, thus indicating
the existence of an ergoregion that extends to infinity.  This also means
that the magnetised Reissner-Nordstr\"om solution is not asymptotic
to the Melvin solution.  Only by setting the charge parameter to zero, so
that the solution reduces to the Schwarzschild-Melvin metric, are these
problems avoided. Section 2 also contains a brief discussion of the
magnetisation of a {\it magnetically} charged Reissner-Nordstr\"om
black hole.  In section 3, we turn to the analysis of the magnetised 
Kerr-Newman solution.  We show that for generic values of the mass, 
charge and rotation parameters $m$, $q$ and $a$ of the original seed 
Kerr-Newman solution, the metric
again necessarily has an ergoregion that extends to infinity, and it is
not asymptotic to the Melvin solution.  We then show that this problem is
avoided if the parameters obey the relation $q=-amB$.  As we discuss in 
detail, the metric {\it is} now asymptotic to the Melvin metric, and we
show how, depending on the choice of time-translation Killing vector, and
the parameters of the solution, there
can be either a compact ergoregion in the neighbourhood of the horizon, or
a toroidal ergoregion outside the black hole, or 
else no ergoregion at all.  In section 4 we make a comparison,
highlighting the similarities and the differences, between the
asymptotically Melvin black holes of this paper and the somewhat analogous
case of black holes in an asymptotically AdS background.  In section 5 we 
discuss the relation between our results for the exact magnetised black
hole solutions and the earlier results of Wald, where the back-reaction
of the external magnetic field on the geometry is neglected.  Our
conclusions, and a discussion of open problems, are in section 6.  In
appendix A we give an explicit dimensional reduction of four-dimensional
Einstein-Maxwell theory to three dimensions, showing how it  gives rise 
to an $SU(2,1)/U(2)$ sigma model coupled to gravity.  We use the $SU(2,1)$
global symmetry in 
appendix B to obtain the complete expressions for the magnetised
Kerr-Newman black hole solution, including both the metric and the vector
potential. Finally, in appendix C, we show how another $SU(2,1)$ 
transformation, applied to a flat space ``seed solution,'' gives rise to
a simple cosmological metric first obtained by Taub, which exhibits some 
features that are rather similar to those we encountered for the
magnetised Kerr-Newman metrics.

\section{Magnetised Reissner-Nordstr\"om Black Hole}

   The general results for the magnetisation of the Kerr-Newman solution are
obtained in appendices A and B.  The expressions are quite complicated, and
so before examining the global properties of the magnetised Kerr-Newman
metrics in section 3, we first specialise to the simpler case where 
the rotation
parameter $a$ is set to zero. 

\subsection{Magnetised electric Reissner-Nordstr\"om} 

  In this section we examine some of the properties of the magnetised
Reissner-Nordstr\"om solution, which can be read off from our results for
the magnetised Kerr-Newman solution in
appendix B by setting the rotation parameter $a$ and the magnetic charge 
parameter $p$ to zero.  After coordinate and gauge transformations
$\phi\rightarrow\phi + 2 m q B^3 \, t$ and $\Phi_0\rightarrow \Phi_0+
3m q B^2$, the solution can be written as 
\bea
d\hat s_4^2 &=& H\, [-f dt^2 + f^{-1}\, dr^2  + r^2 d\theta^2] +
      H^{-1}\, r^2\sin^2\theta\, (d\phi -\omega dt)^2\,,\nn\\
\hat A &=& \Phi_0 dt + \Phi_3 (d\phi-\omega dt)\,,
\eea
where
\bea
f&=& 1- \fft{2m}{r} + \fft{q^2}{r^2}\,,\nn\\
H &=& 1 +\ft12 B^2 (r^2\sin^2\theta + 3 q^2\cos^2\theta) +
  \ft1{16} B^4 (r^2 \sin^2\theta + q^2\cos^2\theta)^2\,,\nn\\
\omega &=& -\fft{2q B}{r} + \ft12 q B^3\, r (1+f \cos^2\theta)\,,\nn\\
\Phi_0 &=& -\fft{q}{r} +\ft34 q B^2 r\, (1+ f\cos^2\theta)\,,\nn\\
\Phi_3 &=& \fft{2}{B} - H^{-1}\Big[\fft{2}{B} + 
             \ft12 B(r^2\sin^2\theta + 3 q^2 \cos^2\theta)\Big]\,.
\eea
The scalar potentials $\psi$, $\chi$ and $\sigma$ arising in the $SU(2,1)$
transformation procedure are given by
\bea
\psi &=& -\fft{q\cos\theta\, 
[1 -\ft14 B^2(r^2\sin^2\theta + q^2\cos^2\theta)]}{H}\,,\nn\\
\chi&=& \fft{2}{B} -\fft1{H}\, \Big[\fft2{B} + \ft12 B
         (r^2\sin^2\theta + 3 q^2\cos^2\theta)\Big]\,,\nn\\
\sigma &=& \fft{q B \cos\theta}{H^2}\, \Big[
q^2 \cos^2\theta -\ft14 B^2(r^2\sin^2\theta + q^2\cos^2\theta) 
  (r^2\sin^2\theta + 4q^2\cos^2\theta)\nn\\
&&\qquad\qquad\quad\quad -
   \ft1{16} B^4 (r^2\sin^2\theta + q^2 \cos^2\theta)^3\Big] \,,
\eea

   The Killing vector
\be
\ell = \fft{\del}{\del t} - \Omega_H\, \fft{\del}{\del\phi}
\ee
becomes null on the horizon at $r=r_+$ where $r_+$ is the larger root of 
$f(r)=0$, and where 
\be
\Omega_H = \fft{2q B}{r_+} -\fft{q r_+ B^3}{2}
\ee
is the angular velocity at the horizon.
The physical electric charge $Q=1/(4\pi) \int {*F}$ is given by
\be
Q = q\, (1-\ft14 q^2 B^2)\,.
\ee

   The magnetised Reissner-Nordstr\"om metric has an
ergoregion where $g_{00}$ becomes positive.  To see this,
we note that
\be
g_{00} = - f H + H^{-1}\omega^2 r^2\sin^2\theta\,.\label{g00}
\ee
Firstly, from the fact that the first term in 
(\ref{g00}) vanishes on
the horizon while the second term contributes positively when 
$\sin\theta\ne 0$, it is evident  
that $g_{00}$ will be positive near to the exterior
of the horizon.  This region is analogous to the ergoregion outside the
horizon of a rotating Kerr black hole.  It can also be seen that $g_{00}$
will be positive
close to the polar axes at large $r$ with $\sin\theta$ becoming small
such that
$r\sin\theta$ is held fixed.  To see this, it is convenient to introduce 
cylindrical  coordinates
$\rho$ and $z$, defined by
\be
\rho=r\sin\theta\,,\qquad z=r\cos\theta\,.\label{weylcoords}
\ee
Making an expansion of $g_{00}$ in inverse powers of $z$,
we find 
\be
g_{00} = \fft{16 B^6 q^2 (z^2+ 2m z)\rho^2 }{
   16+ 8 B^2 (\rho^2+3 q^2) + B^4(\rho^2+q^2)^2} +
   {\cal O}(z^0)\,,
\ee
and therefore 
$g_{00}$ can be arbitrarily large and positive at large $z$, holding
$\rho$ fixed.
A numerical study of the metric function $g_{00}$ reveals that the two
regions of positivity described above are in fact connected.  A plot
showing the ergoregion for a representative example is shown in figure 1 
below.  The ergoregion extends to infinity near the poles regardless of how
small $B$ or $q$ are, provided that they are non-zero.  It asymptotically
approaches the $z$ axis as $z$ tends to infinity. 

\begin{figure*}[h!]
\centering
\includegraphics[height=60mm]{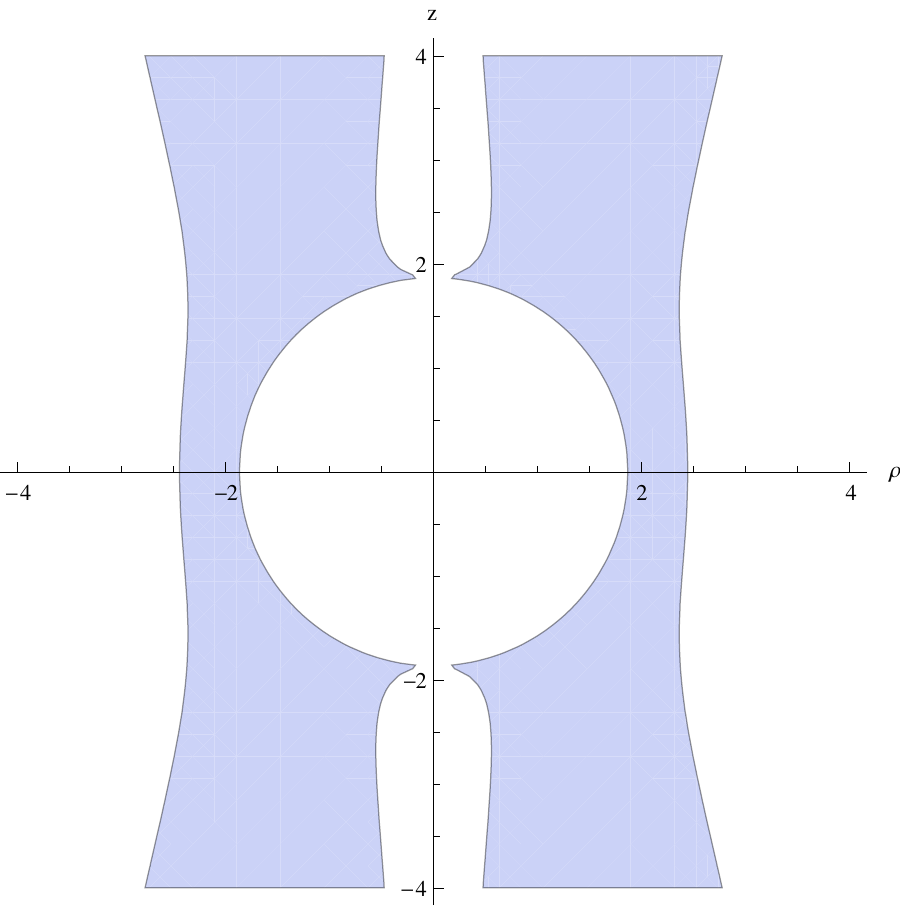}
\caption{\it The shading indicates a cross-section through 
the ergoregion for a magnetised 
Reissner-Nordstr\"om black hole, with $M=1$, $B=1$, $q=\ft12$.  The full 
ergoregion is the surface of revolution obtained by rotating this around
the vertical axis.  The ergoregion 
extends to infinity in the positive and negative $z$ directions.}
\end{figure*}

Note that the ergoregion is absent if $q=0$, in which case the metric
reduces to the Schwarzschild-Melvin solution.  

Although simpler than the
general magnetised Kerr-Newman case, the magnetised Reissner-Nordstr\"om metric
is still quite complicated.  Near infinity, it resembles a much simpler,
but little known, stationary vacuum metric, which is described in appendix C.
That metric also exhibits an ergoregion near infinity, which is qualitatively
similar to the more complicated ergoregion in the magnetised black holes. It
can be obtained by starting from flat space and then acting with 
an $SU(2,1)$ transformation within the class described
in appendix A.

\subsection{Magnetised magnetic Reissner-Nordstr\"om}

   For completeness, we record here the expressions for the magnetised
Reissner-Nordstr\"om solution carrying a magnetic, rather than electric, 
charge. This solution is obtained by setting $a=0$ and $q=0$ in the results
for the magnetised Kerr-Newman solution that are obtained in appendix B.
\bea
ds_4^2 &=& H\, [ -f dt^2 + f^{-1}\, dr^2 + r^2 d\theta^2] + H^{-1}\, 
  r^2\sin^2\theta\, d\phi^2\,,\nn\\
A&=& \Phi_3\,d\phi\,,
\eea
with
\bea
f&=& 1 -\fft{2m}{r} +\fft{p^2}{r^2}\,,\nn\\
H &=& [1 +\ft14 B^2 r^2 \sin^2\theta - p B \cos\theta + 
                \ft14 p^2 B^2 \cos^2\theta]^2\,,\nn\\
\Phi_3 &=& \fft{-p\cos\theta + \ft12 B(r^2\sin^2\theta+p^2\cos^2\theta)}{
 \sqrt{H}}\,.
\eea

\section{Properties of the Magnetised Kerr-Newman Black Hole}

   In this section, we investigate some of the properties of the magnetised
Kerr-Newman solution, which is constructed in appendix B using the appropriate
$SU(2,1)$ transformations described in appendix A.  For simplicity, we shall
restrict attention to the case where the original seed solution is a
Kerr-Newman solution carrying purely electric charge.  Thus, we set $p=0$ in
all the results obtained in appendix B.

\subsection{Electric and magnetic charges}

   If we set $p=0$ in the magnetised Kerr-Newman solution, then
the requirement of no conical deficit at the poles of the sphere 
implies that the azimuthal angle $\phi$ should have period \cite{Hiscock}
\bea
\Delta\phi &=& 2\pi H|_{\theta=0} =2\pi H|_{\theta=\pi}\nn\\
&=& 2\pi\Big[1 +\ft32 q^2 B^2 + 2 a q m B^3 +(a^2 m^2 +\ft1{16} q^4) B^4\Big]
\,.
\eea
Thus the conserved electric charge is given by
\be
Q= \fft1{4\pi}\int_{S^2} {{\hat *}\hat F} = 
    \fft{\Delta\phi}{4\pi} \Big[ \psi\Big]_{\theta=0}^{\theta=\pi}\,,
\ee
and hence
\be
Q= q(1-\ft14 q^2 B^2) + 2 a m B\,.\label{Qcharge}
\ee
Note that
to obtain a neutral black hole with conserved charge $Q=0$, we need to 
start with a charged rotating  black hole with charge given by
$q(1-\frac{1}{4}q^2  B^2 )  = - 2am B$. In the limit that $qB <<1$ we 
recover Wald's result (\ref{wald}).

The conserved magnetic charge $P=1/(4\pi) \int \hat F$ is equal to zero in 
this case where we set $p=0$.

\subsection{Ergoregions}

As in the Kerr or Kerr-Newman solution itself 
we expect, of course, that there should be a compact  
ergoregion in the vicinity of the exterior of the horizon.  However, 
in the magnetised Kerr-Newman solution the ergoregion in general 
extends out to infinity in the vicinity of 
the rotation axis.  As we shall discuss, there is one exceptional
circumstance where this does not occur, and that is if
the charge parameter in the solution is chosen to be given by $q=-a m B$.  
If we substitute this relation into the expression (\ref{Qcharge}) for the
electric charge $Q$ on the black hole, we find
\be 
Q= amB(1+\ft14 a^2 m^2  B^4)\,.\label{physQ}
\ee
It is striking that in the small-$B$ limit the magnitude of the charge is
one half that obtained by Wald (\ref{wald}).

   Consider the Killing vector field
\be
K_\Omega = \fft{\del}{\del t} + \Omega\, \fft{\del}{\del \phi}\,,
\label{Kdef}
\ee
where the angular velocity $\Omega$ is a constant which we shall choose later.
   If we look at large distances while holding the polar angle $\theta$
fixed, the dominant term in the large-$r$ expansion of 
$K_\Omega^\mu K_\Omega^\nu \, g_{\mu\nu}$ is negative, and 
has the generic form that one expects in a Melvin
universe, with
\be
K_\Omega^\mu K_\Omega^\nu \, g_{\mu\nu} 
= -\ft1{16} B^4 r^4 \sin^4\theta + {\cal O}(r^3)\,.
\ee
Thus the Killing vector $K_\Omega$ is timelike at large $r$, 
for fixed $\theta$, for any choice of
$\Omega$.
However, if we look at the region where $r$ is large but instead $r\sin\theta$
is held fixed, then it turns out that $K_\Omega$  becomes spacelike,
again for any value of $\Omega$, signaling the occurrence of an ergoregion
that extends out to infinity near the rotation axis.
To see this, it is convenient to 
use cylindrical  coordinates
$\rho$ and $z$, as defined in (\ref{weylcoords}), instead of $r$ and $\theta$.  
We are then interested in probing the region 
where $z$ is large while $\rho$ remains small.

With $p$ taken to be zero,
we find that the expansion of 
$K_\Omega^\mu K_\Omega^\nu \, g_{\mu\nu} $ in
inverse powers of $z$ is given by
\be
 K_\Omega^\mu K_\Omega^\nu \, g_{\mu\nu}= 
 \fft{16  B^6(q+a m B)^2 \rho^2 }{W} \, z^2 -
   \fft{4 B^6 (q+a m B) [8q m + aB(q^2+4m^2)]\rho^2 }{W}\, z 
+ {\cal O}(z^0)\,,
\ee
where $W$ is the positive quantity
\be
W= 16+ 8 B^2\rho^2
   + B^4(\rho^2 + q^2)^2 + 24B^2(q+\ft23 a m B)^2 +
       \ft{16}{3} a^2 m^2 B^2\,.
\ee
Thus $K_\Omega^\mu  
K_\Omega^\nu \, g_{\mu\nu}$ will become large
and positive in this region unless we choose $q= -a m B$.

   Whilst it is easy to see that there is an ergoregion near the rotation axis
that extends out to infinity if $q\ne - a m B$, more work is required 
to establish what happens 
if $q= -amB$. As a start, we may investigate the 
large-$z$ region at fixed $\rho$, where we saw that $K_\Omega$ 
became spacelike in the previous discussion.
Setting
$q=-amB$ and expanding in inverse powers of $z$, we now find
\be
 K_\Omega^\mu K_\Omega^\nu g_{\mu\nu} = -\,
\fft{F_+\, F_-}{16(4 + a^2 m^2 B^4 + B^2 \rho^2)^2} + {\cal O}(z^{-1})
\,,\label{FpFm}
\ee
where
\bea
F_\pm &=&(4+ B^2\rho^2)^2 + 2 a^2 m^2 B^4(4+ B^2\rho^2) 
 + a^4 m^4 B^8\nn\\
&& \pm [16\Omega +2a m^2 B^4 (12+a^2 B^2)] \rho\,.\label{Fpmdef}
\eea
By choosing the angular velocity to be given by
\be
\Omega =\Omega_s\equiv  -\ft18 a m^2 B^4(12 + a^2 B^2)\,,\label{Omegasol}
\ee
we have
\be
F_+=F_-= (4+ B^2 \rho^2 + a^2 m^2 B^4)^2\,,
\ee
and thus 
\be
 K_\Omega^\mu K_\Omega^\nu g_{\mu\nu} = -\ft1{16}
 (4 + a^2 m^2 B^4 + B^2 \rho^2)^2 + {\cal O}(z^{-1})
\,,\label{KK}
\ee
We see that with the angular velocity $\Omega$ chosen as in 
(\ref{Omegasol}), the Killing vector $K_\Omega$ 
defined in (\ref{Kdef})
is timelike in this region.  Thus it appears that when $q=-amB$, this Killing
vector $K_\Omega$ is timelike everywhere at large distances, and so
the ergoregion is now confined to the neighbourhood of the horizon.

   Further insight into the significance of the angular velocity $\Omega_s$
defined in (\ref{Omegasol}) can be obtained by introducing a comoving 
coordinate 
\be
\tilde\phi = \phi-\Omega\, t\,.\label{tildephidef}
\ee
An examination of the metric component $g_{t\tilde\phi}$ at large $z$ and
small $\rho$ reveals that unless $q=-am B$, it diverges linearly with $z$.
If one restricts to $q=-amB$, one finds
\be
g_{t\tilde\phi}= \fft{2(8\Omega +12 a m^2 B^4 + a^3 m^2 B^6)\rho^2}{
 (4+ a^2 m^2 B^4 + B^2 \rho^2)^2} + {\cal O}(\fft1{z})\,.
\ee
Evidently, if one makes the choice $\Omega=\Omega_s$, then the cross term
$g_{t\tilde\phi}$ vanishes to lowest order in $\rho$, at large $z$.  In fact,
the large $z$ expansion now takes the form
\be
g_{t\tilde\phi}= -\fft{8 a m B^2(4+a^2 m^2 B^4)\rho^2}{
(4+ a^2 m^2 B^4 + B^2 \rho^2)^2\, z} + {\cal O}(\fft1{z^2})\,.
\ee
The significance of the absence of a term in $\rho^2$ at large $z$ is that
the Killing vector field $K_{\Omega_s}$ is locally static, that is,
the ``twist vector''
\be
\tilde\omega_\Omega^\mu= \epsilon^\mu{}_{\nu\rho\sigma}\, K^\nu_\Omega\,
\nabla^\rho K^\sigma_\Omega
\ee
vanishes on the axis if we choose $\Omega=\Omega_s$.  Thus, the choice
of azimuthal coordinate $\tilde\phi$ given by (\ref{tildephidef}), 
with $\Omega=\Omega_s$,
 provides the best approximation to a locally non-rotating
inertial frame near the axis.

  It can be easily verified that $|K_{\Omega_s}|^2$ is negative everywhere
at large distances, and that outside the horizon, it becomes positive
only within a compact ergoregion in the neighbourhood of the horizon.  This
ergoregion can be thought of as a deformation, induced by the external 
magnetic field, of the usual ergoregion outside a Kerr or Kerr-Newman
black hole.

   As we shall discuss in detail in the next section, it is in fact
possible to find a different choice of Killing field $K_\Omega$ that is,
for sufficiently large $B$,
timelike everywhere outside the horizon.  Specifically, we do this
by taking $\Omega=\Omega_H$, the angular velocity of the horizon. This
is defined by the condition that $K_{\Omega_H}$ be null on the horizon.  
Bearing in mind that we are setting $q=-amB$ in the Kerr-Newman seed
solution, we see from (\ref{knfns}) that the horizons of the magnetised black
hole are located at the roots of
\be
r^2 -2m r + a^2(1+m^2 B^2) =0\,.
\ee
It is therefore convenient then to express the rotation parameter $a$
as a fraction
$\varepsilon$ of the maximum (extremal) value:
\be
a = \fft{\vep\, m}{\sqrt{1+m^2 B^2}}\,,\qquad 0\le \varepsilon\le 1\,.
\label{apara}
\ee
Defining then the ``co-extremality parameter'' $\tilde\vep$ by $\vep^2=
1-\tilde\vep^2$, we see that the outer and inner horizons are located at
\be
r_\pm = (1\pm \tilde\vep) \, m\,.
\ee
We then find that $\Omega_H$ is given by
\be
\Omega_H =
\fft{(1-\tvep)[8+ 4(7+3\tvep)m^2 B^2 -2(1+6\tvep+\tvep^2)
m^4 B^4 -(1+\tvep)(21+2\tvep+\tvep^2) m^6 B^6]}{16m (1-\tvep^2)^{1/2}
(1+m^2 B^2)^{3/2}}\,.\label{OmegaH}
\ee
If $B$ exceeds a certain value $B_+$, which can be determined
as the smallest positive root of a rather complicated 18th-order polynomial
in $B^2$ that we shall not present here, then numerical studies 
indicate that $K_{\Omega_H}$
is timelike everywhere outside the horizon.  If $B$ is taken to lie in
a range $B_-<B<B_+$, then an ergoregion of toroidal topology develops outside,
and disjoint from, the horizon, in the equatorial plane.\footnote{Numerical
studies indicate that as $B$ approaches $B_+$ from below, the toroidal 
ergoregion contracts down to a ``Saturn ring'' in the equatorial plane, which
finally disappears when $B$ reaches $B_+$.
Hence $B_+$ can be determined from the requirement that $H(r,\theta)\equiv
|K_{\Omega_H}|^2$ and $\del H(r,\theta)/\del r$, evaluated in the equatorial
plane $\theta=\pi/2$, should simultaneously vanish.}  As $B$ approaches
$B_-$ from above, the toroidal ergoregion develops ``lobes'' that extend
further and further upwards and downwards along the $z$ direction.  If $B$
is smaller than $B_-$, these lobes extend all the way to infinity.  The 
value of $B_-$ is determined as the smallest positive root of the 
polynomial
\bea
&&256(1-\tvep)(1+\tvep)^2\, m^6 B^6 + (349-889\tvep+567\tvep^2 + 243\tvep^3)\,
m^4 B^4 \nn\\
&&+ 4(121+310\tvep+ 81\tvep^2)\, m^2 B^2 -108(1-\tvep)=0\,.
\eea
This polynomial is determined by looking at the leading-order term in
the large-$z$ expansion of $|K_{\Omega_H}|^2$ expressed in the cylindrical
coordinates $\rho$ and $z$.  

  In figure 2 below, we present plots showing a cross-section of 
the location of the ergoregion
for $K_{\Omega_H}$ for three representative choices of the parameters.  
(The full ergoregion is the surface of revolution obtained by rotating the
plot around the vertical axis.)
In the first, $B$ is less than $B_-$ and the ergoregion extends to 
infinity.  In the second, $B$ lies between $B_-$ and $B_+$, and
so there is a toroidal ergoregion, disjoint from the horizon. In the 
third plot, $B$ is larger but still less than $B_+$, and so the toroidal
ergoregion has contracted.

\begin{figure*}[h!]
\centering
\mbox{\subfigure{
\includegraphics[height=50mm]{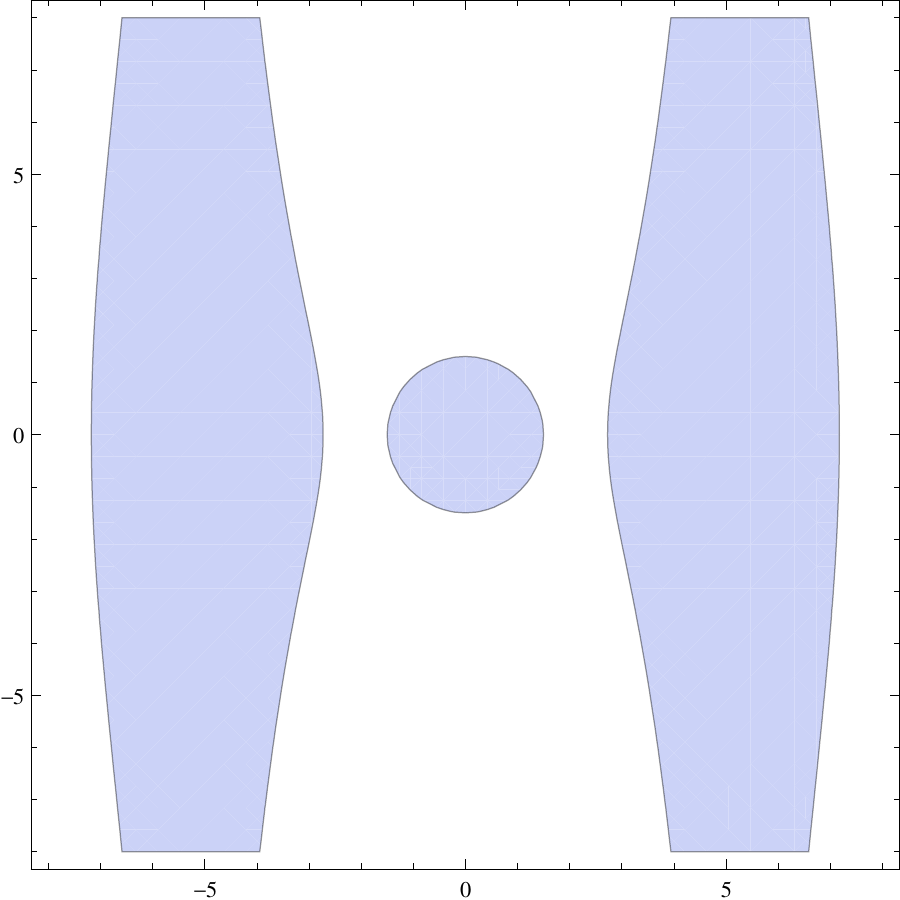}}\quad
\subfigure{
\includegraphics[height=50mm]{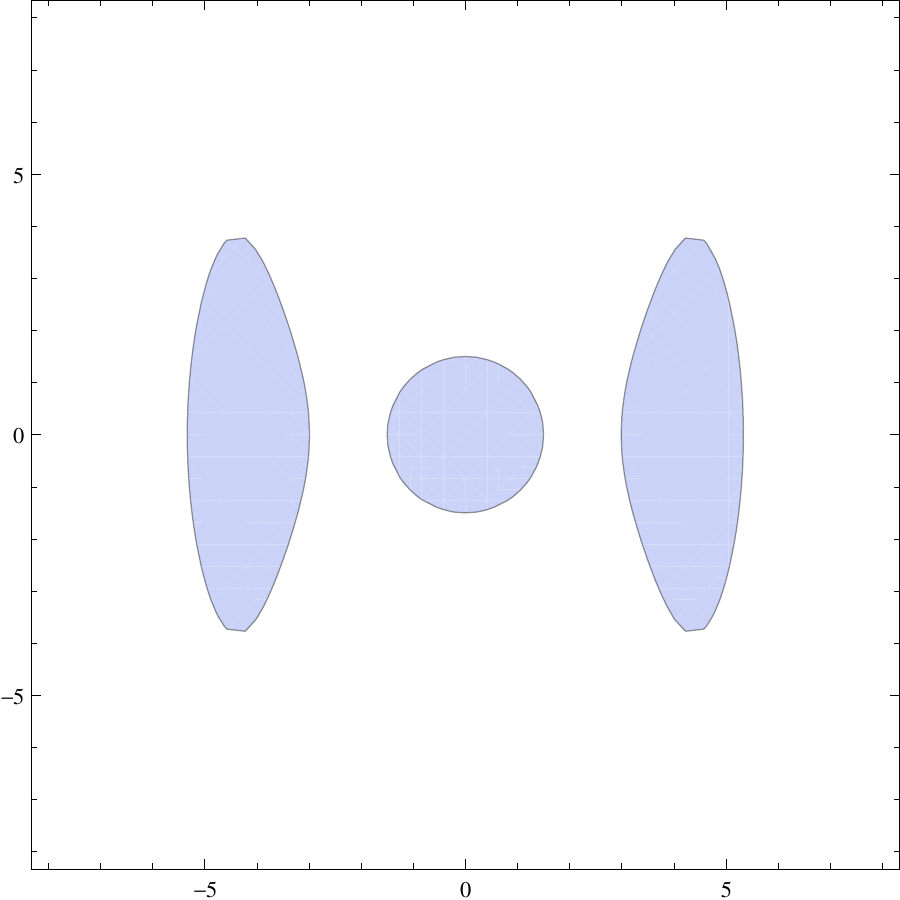}}\quad
\subfigure{
\includegraphics[height=50mm]{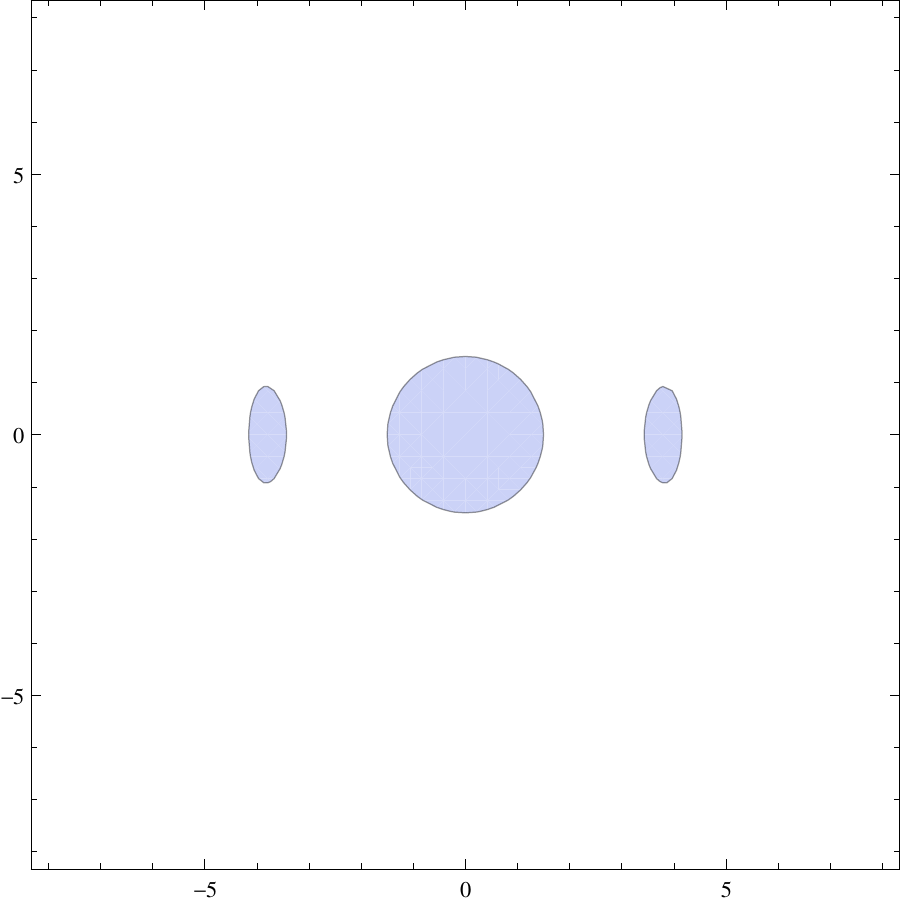}}
}
\caption{\it The ergoregion (shaded) for the Killing vector $K_{\Omega_H}$
for three example parameter choices. In the first,
$B <B_-$, and the ergoregion outside the horizon extends to infinity.  In
the second, $B_-<B<B_+$, and the ergoregion is a torus outside the horizon.
The third plot is for a larger value of $B<B_+$, showing how the torus is 
contracting.  (The interior of the horizon is shown shaded here too, so that
its location relative to the ergoregion is evident.)}
\end{figure*}

   It is perhaps worthwhile also to comment on what happens if one simply
takes the ``naive'' choice $\Omega=0$ in the definition of the time-translation
Killing vector $K_\Omega$.  In other words, if one simply uses the original
$t$ and $\phi$ coordinates of the Kerr-Newman seed solution as the time
and azimuthal angle.  For sufficiently small values of the magnetic field
$B$, the Killing vector $\del/\del t$ is timelike everywhere outside the
black hole except for a Kerr-Newman-like ergoregion near the horizon.
(As usual, it is to be understood in this discussion that we are taking
$q=-amB$.)  As $B$ is increased, a value $B=B_{\rm crit}$ is reached for which
the ergoregion disappears altogether.  This corresponds to the value of $B$ at
which the angular velocity $\Omega_H$ given in (\ref{OmegaH}) vanishes.  If
$B$ is increased beyond $B_{\rm crit}$, the ergoregion develops again around
the horizon, and begins to grow ``lobes'' that extend upwards and downwards 
close to the axis of rotation.  Eventually, if the magnetic field reaches or
exceeds a certain value $B_{\rm max}$, these lobes extend all the way to 
infinity close to the rotation axis.  The value of $B_{\rm max}$ is determined
by the condition that there exist a $\rho$ such that $F_-(\rho)$ defined in
(\ref{Fpmdef}) satisfy $F_-(\rho)=0$ and $dF_-(\rho)/d\rho=0$ 
simultaneously (with $\Omega=0$).  This determines that $B_{\rm max}$ is
the smallest positive root of 
\bea
&&64 \vep^6 m^{12} B^{12} - 
 3\vep^2(36-16\vep+3\vep^2)(36+16\vep+3\vep^2)m^{10} B^{10} -
 24\vep^2(196-5\vep^2) m^8 B^8 \nn\\
&&
+ 16(256+141 \vep^2)m^6 B^6 +
3072 (4+\vep^2) m^4 B^4 + 12288 m^2 B^2 + 4096=0\,.\label{Bmaxpoly}
\eea
We present plots below, in figure 3, showing the ergoregion for the Killing 
vector $\del/\del t$ for two representative examples.  In the first, 
we have $B_{\rm crit} < B < B_{\rm max}$ and prominent lobes are visible
in the neighbourhood of the horizon.  In the second, we have $B>B_{\rm max}$, 
and the lobes extend out to infinity.

\begin{figure*}[h!]
\centering
\mbox{\subfigure{
\includegraphics[height=60mm]{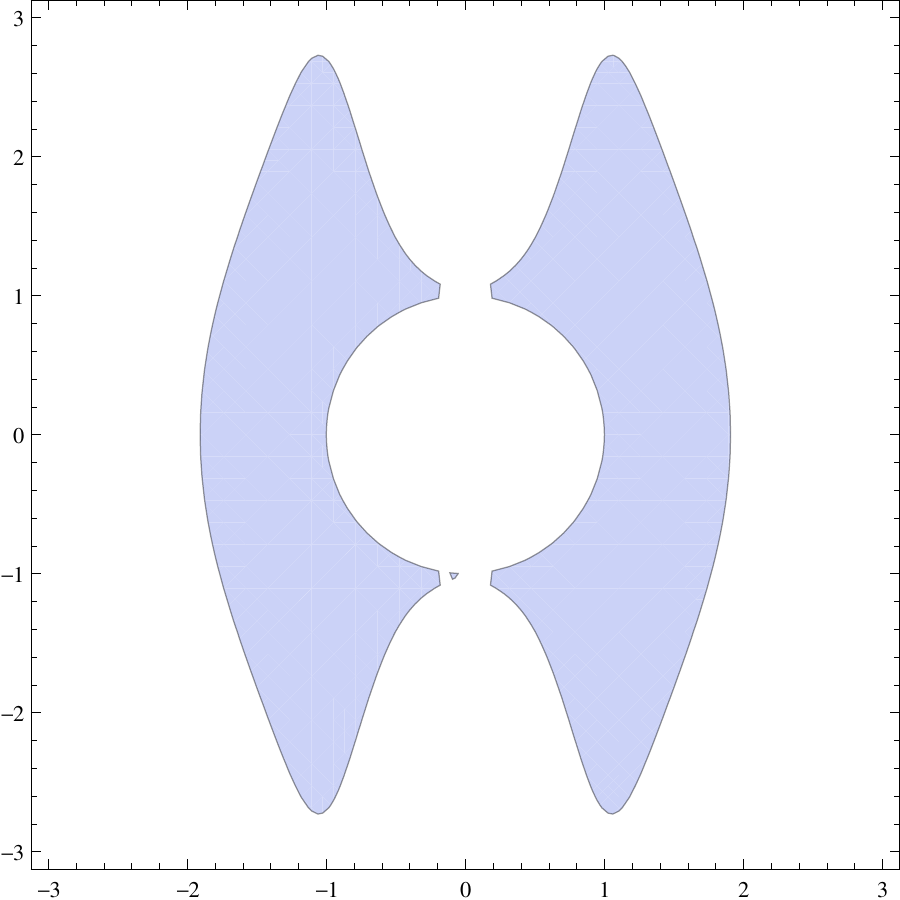}}\quad
\subfigure{
\includegraphics[height=60mm]{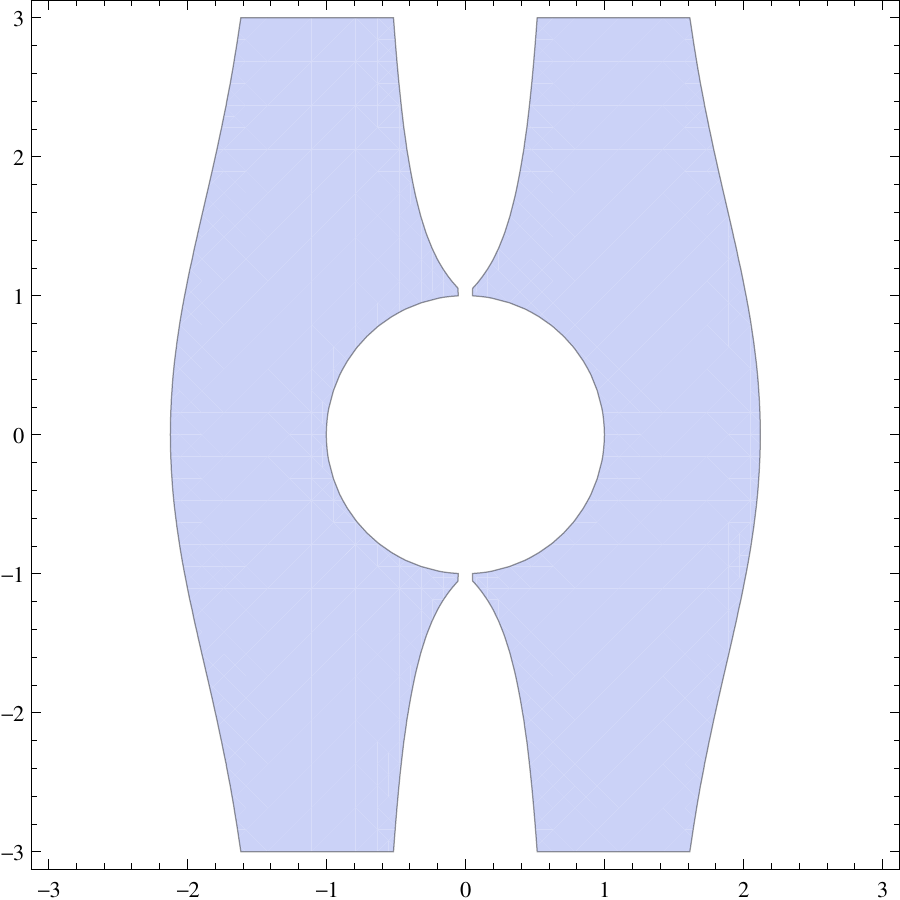}}
}
\caption{\it The ergoregion (shaded) for the Killing vector $\del/\del t$
for two example parameter choices. In the first, 
$B_{\rm crit} < B <B_{\rm max}$ and prominent lobes have developed. In
the second, $B>B_{\rm max}$ and the lobes extend to infinity.}
\end{figure*}

\subsection{The asymptotic structure of the solution}

   We saw in section 3.2 that if we make the choice $q=-amB$, and introduce
the new azimuthal coordinate $\tilde\phi=\phi -\Omega_s\, t$, then the metric
that we have constructed is asymptotic to the static Melvin metric.  We
now turn to a consideration of the electromagnetic field.

Using the
expression (\ref{vierbeinEH}) for the vierbein components of the magnetic
and electric fields, we find that close to the axis at large $z$, the
radial components take the form
\bea
H_r = &=& \fft{B}{(1+ \ft14 a^2 m^2 B^4 + \ft14 B^2\rho^2)^2} +
{\cal O}(\fft1{z}) \,,\nn\\
E_r = &=& \fft{-\ft12 a m B^3}{(1+ \ft14 a^2 m^2 B^4 + \ft14 B^2\rho^2)^2} +
{\cal O}(\fft1{z})\,,
\eea
and so there is not only a constant magnetic field along the $z$ axis but
also a constant electric field too.\footnote{An analogous observation
about the presence of an asymptotically constant electric field along the
$z$ axis was made previously in \cite{Hiscock}.  We are grateful to
Marco Astorino for drawing this to our attention.}
The reason why there is also an asymptotically constant electric field
is presumably because the electric flux originating from the black hole
does not spread out as it would in the absence of the applied magnetic
field, but instead it is confined to the neighbourhood of the axis by
the Melvin geometry.
  
In other words, the solution is
asymptotic to a certain duality rotation of the Melvin magnetic universe.
We may therefore apply a counter duality transformation to our solution, chosen
so as to make it asymptotic to the purely magnetic Melvin universe.  Thus
we consider the transformation
\be
H_r' = H_r\, \cos\alpha + E_r\, \sin\alpha,\qquad
E_r'= E_r\, \cos\alpha - H_r\, \sin\alpha\label{dual1}
\ee
with $\tan\alpha= -\ft12 a m B^2$, leading to the asymptotic forms
\bea
H_r' = &=& \fft{B\, \sqrt{1+\ft14 a^2 m^2 B^4}}{
 (1+ \ft14 a^2 m^2 B^4 + \ft14 B^2\rho^2)^2} +
{\cal O}(\fft1{z}) \,,\nn\\
E_r' = &=&  {\cal O}(\fft1{z})\,,
\eea

   The duality transformation (\ref{dual1}) 
will also rotate the physical conserved
electric and magnetic charges $P$ and $Q$.  We previously had $P=0$, and
$Q$ given by (\ref{physQ}).  After the rotation, we shall have
\be
Q'= a m B \, \sqrt{1+\ft14 a^2 m^2 B^4}\,,\qquad
P'= -\ft12 a^2 m^2 B^3\,\sqrt{1+\ft14 a^2 m^2 B^4}\,.
\ee
From a physical point of view, however, it is perhaps more natural to
maintain the original choice of duality complexion, so that the black
hole carries only electric charge and the asymptotically Melvin background
has electric as well as magnetic flux.

\section{Comparison with Kerr-AdS Spacetime}

  In order to gain some intuition for what is happening, we shall recall some
facts about rigidly rotating reference 
systems in asymptotically flat and asymptotically
anti-de Sitter spacetimes.  In flat spacetime it has been understood
since early discussions
of Born rigidity \cite{Born,lang1,lang2}, 
the Ehrenfest paradox \cite{Ehrenfest}   
and the  Sagnac effect 
\cite{Sagnac} that if one passes to a rigidly-rotating coordinate system,
then it cannot be extended beyond the ``velocity of light cylinder'' situated
at $\rho=\Omega^{-1}$, beyond which the co-rotating Killing vector 
$K_\Omega=\del/\del t + \Omega \del/\del\phi$ becomes spacelike.
Following \cite{lang1,lang2} we introduce a coordinate 
$\tilde \phi= \phi -\Omega t$, where $t,\phi,\rho,z$ are 
cylindrical inertial coordinates for Minkowski spacetime,  
in which the flat metric takes the Langevin form
\ben
ds^2  = -(1- \Omega ^2 \rho  ^2) 
\bigl (dt - \Omega  {\rho ^2 d {\tilde \phi}  \over 1-
  \Omega ^2 \rho ^2 } \bigr ) ^2 +dz ^2 + d \rho ^2 + 
{ \rho ^2  \over 1-  \Omega ^2 \rho ^2 } d {\tilde \phi} ^2 
\,. \label{langevin} 
\een

Note that $\tilde \phi$ is constant along the orbits of $K_{\Omega}$, i.e.\
$K_{\Omega} \tilde \phi =0$, and the 3-metric
\ben
ds^2_\perp = dz ^2 + d \rho ^2 + 
{ \rho ^2  \over 1-  \Omega ^2 \rho ^2 } d {\tilde \phi}^2 
\label{orthog}\,, 
\een
orthogonal to the orbits, is independent of time. Thus the 
coordinates $t,\tilde \phi, z, \rho$ are rigidly rotating
in the sense of Born \cite{Born}. Because the twist 1-form
\ben
 \star (K_ \Omega)_\flat \wedge d  (K_ \Omega)_\flat = 2 \Omega dz   
\een
is non-vanishing,  there is no hypersurface orthogonal to the orbits
of $K_\Omega$, and the curved metric $ ds^2_\perp$ is not
the induced metric on any such surface. This resolves
Ehrenfest's paradox \cite{Ehrenfest}. The cross term in the metric , that is
the term $2 \Omega \rho ^2 dt d \tilde \phi $, 
gives the Sagnac effect \cite{lang1,lang2}. 

For a general stationary axisymmetric spacetime
with adapted coordinates $ t,\phi, x^A$, with $A=1,2$, we have
\ben
ds ^2 = - e^{2U(x^A)} (dt + \omega(x^A)  d \phi) ^2 + 
 g_{AB}(x^A)  dx^A dx ^B  + X(x^A)  d \phi ^2  
\,.
\een
The 1-form  $\omega d \phi$ is the Sagnac connection \cite{Ashtekar},
and if its curvature $d\omega \wedge d \phi$ is non-zero
the Killing field $K=\frac{\p}{\p t}$ is locally rotating.
Changing coordinates by setting $\tilde \phi = \phi-\Omega t$, where
$\Omega$ is constant, gives a new metric for which
\ben
g_{tt} = -e^{2 \tilde U} = -e^{2U} (1+ \omega \Omega)^2 + X \Omega ^2 
\een 
and 
\ben
g_{t\tilde \phi}= -e^{2 \tilde U} \tilde \omega
= -e^{2 U} (1 +\omega \Omega) \omega + X \Omega 
\,.
\een    
One cannot expect in general to be able to eliminate
$\omega$ by  this means, but it may be possible to 
make the Sagnac curvature $d \omega$ vanish along one orbit
of $K_\omega$. If so, this choice $\Omega=\Omega _s$  will define 
a locally static reference frame on the orbit.
Of course, passing to  new rigidly rotating reference system
will mean  that the domain of strict stationarity for which
$g_{tt}$ is negative will change.  Thus, for example,  a
frame rotating with the angular velocity of a Kerr black hole breaks down
outside an analogous velocity of light surface.  Particles with 
future-directed timelike momenta $p_\mu$ outside the velocity of light surface
may carry negative energy with respect to the co-rotating Killing
vector $K_\Omega$, i.e. $-p_\mu K^\mu_\Omega <0$. Thus from the 
point of view of a co-rotating observer, the region where 
$K_\Omega$ is spacelike is potentially a source of energy, i.e.\
it is an ergoregion.  Moreover, every rotating observer has such an 
ergoregion.   On the other hand, observers who are not rotating at
infinity will find an ergoregion surrounding the black hole. In other
words, the concept of an ergoregion, and its location, is observer dependent.
However, there is no choice of Killing vector field that is timelike
everywhere outside the horizon of a Kerr black hole, and so any observer
will see an ergoregion somewhere in the exterior spacetime.

   This need not, however, be the case for asymptotically anti-de Sitter (AdS)
spacetimes.  In AdS itself, it is possible to pass to a rotating frame in
which the Killing vector $K_\Omega= \del/\del t + \Omega \del/\del\phi$ 
is timelike everywhere, as long as 
\be
\Omega^2 < \ell^{-2}\,,
\ee
where $\ell$ is the AdS radius. For this reason, when dealing with
asymptotically AdS spacetimes, we need an extra criterion
to decide whether or not we are in a frame that is ``non-rotating at 
infinity.''  This can be done by requiring that the conformal boundary metric
be non-rotating.  For the Kerr-AdS black hole, the metric in this frame is
given by
\be
ds^2= -\fft{\Delta_\theta\, X}{\Xi^2\, R^2}\, 
\Big(dt + \fft{2 a m r \sin^2\theta}{X}\, d\phi\Big)^2 +
  \fft{\Delta_r\, R^2\sin^2\theta}{X}\, d\phi^2 
+R^2\, \Big(\fft{dr^2}{\Delta_r} +\fft{d\theta^2}{\Delta_\theta}\Big)\,,
\ee
where
\bea
\Delta_r &=& (r^2+a^2)(1+r^2\, \ell^{-2}) - 2mr \,,\qquad
\Delta_\theta= 1 - a^2\, \ell^{-2}\, \cos^2\theta\,,\\
X &=& \Xi\, (1+r^2\, \ell^{-2})\, R^2 - 2mr \Delta_\theta\,,\qquad
R^2=r^2+a^2\cos^2\theta\,,\qquad \Xi=1-a^2\, \ell^{-2}\,.\nn
\eea
The importance of this non-rotating frame 
is that questions of energy, stability and black-hole thermodynamics
become much simpler and better defined \cite{GPP}.
 
  In dealing with rotating black holes in anti-de Sitter backgrounds, we 
could of course pass to a frame that is co-rotating with respect to the black
hole.  One may ask whether such a frame may be extended all the way to 
infinity, or whether it has a velocity of light surface beyond which a 
co-rotating Killing vector 
\be
\widetilde K_H\equiv 
  \fft{\del}{\del t} + \Omega_H \, \fft{\del}{\del\phi}\,,\qquad
\Omega_H = \fft{a(1+r_+^2\, \ell^{-2})}{(r_+^2+a^2)}
\ee
becomes spacelike. 
If the black hole is sufficiently small that $r_+^4 < a^2 \ell^2$, there is
a velocity of light surface analogous to that in asymptotically flat 
spacetimes.  On the other hand, for black holes such that
$r_+^4 > a^2 \ell^2$, the Killing vector $\widetilde K_H$ is timelike
everywhere outside the horizon.  Thus in contrast to the asymptotically
flat case, for sufficiently large black holes there is a choice of Killing
vector field that is timelike everywhere outside the horizon, and
thus it has no ergoregion.

   The situation in the case of the magnetised black holes that we are 
considering in this paper is more involved.  It is helpful to consider 
first the Melvin universe without a black hole.  In static coordinates,
the metric is
\be
ds^2 = (1+\ft14 B^2\rho^2)^2(-dt^2+d\rho^2+ dz^2) + 
\fft{\rho^2 d\phi^2}{(1+\ft14 B^2\rho^2)^2}\,.
\ee
Introducing the rotating coordinate $\tilde\phi$ as in (\ref{tildephidef}),
which is constant along the orbits of the Killing field $K_\Omega
=\del/\del t+ \Omega\del/\del\phi$, i.e. $K_\Omega (\tilde\phi)=0$,
the metric becomes
\be
ds^2 = (1+\ft14 B^2\rho^2)^2(-dt^2+d\rho^2+ dz^2) + 
\fft{\rho^2 (d\tilde\phi +\Omega dt)^2 }{(1+\ft14 B^2\rho^2)^2}\,.
\ee
We note that the metric component $g_{t\tilde\phi}$ is non-vanishing,
and proportional to $\rho^2$ when $\rho$ is small, and that 
\be
g_{tt} = -(1+\ft14 B^2\rho^2)^2\, \Big(1-
 \fft{\Omega^2\rho^2}{(1+\ft14 B^2\rho^2)^4}\Big)\,.
\ee
From this it can be seen that if $\Omega^2<(4/3)^3\, 
B^2$ then the rigidly rotating
Killing vector $K_\Omega=\del/\del t+ \Omega\del/\del\phi$
is everywhere timelike.  If $\Omega^2> (4/3)^3\, B^2$, it becomes 
spacelike within the
annular cylinder $0<\rho_- <\rho<\rho_+$, with $\rho_- < (1/\sqrt3)\, 
\rho_{\rm Melvin}
<\rho_+$, where $\rho_{\rm Melvin}$ is the Melvin radius, defined by
\be
\rho_{\rm Melvin} =\fft{2}{B}\,.\label{melvinradius}
\ee
This is an indication that a general magnetised black hole solution could,
for a small enough $B$ field,
be expected to have an ergoregion within an annular cylinder
extending to infinity unless the coordinate system is chosen to be
asymptotically static.

  Turning now to the magnetised Kerr-Newman solutions, we have seen 
that in the general case $q\ne -amB$ the situation is much more pathological
than in the Melvin universe example we have just been considering. 
Namely, there is {\it no} choice of Killing vector field,
i.e. no choice of $\Omega$ in the definition (\ref{Kdef}),
that does not have an ergoregion in the neighbourhood of infinity.  

   In the 
special case when $q=-amB$, then for given $a$, $m$ and $B$, or equivalently
$a$, $r_+$ and $B$, there is a range of values for $\Omega$ such that
$K_\Omega =\del/\del t +\Omega \del/\del \phi$ is timelike
at infinity.  This range includes $\Omega=\Omega_s$, defined in 
(\ref{Omegasol}),
for which $K_\Omega$ is timelike at infinity for all values of $a$, $m$ 
and $B$. 
There is also a different choice of $\Omega$ within
this range, namely the angular velocity of the horizon, 
$\Omega=\Omega_H(a,m,B)$, for which, provided that $B$ is sufficiently 
large ($B>B_+$, 
defined in section 3.2), 
$K_\Omega$ is timelike {\it everywhere} outside the
horizon (and lightlike {\it on} the horizon).  
This situation is analogous to what we saw in the case of AdS black holes.
If, however, $B$ is less than $B_+$ then for a range $B_-<B<B_+$ (with $B_-$
defined in section 3.2), there is a toroidal ergoregion outside and disjoint
from the horizon.  If $B\le B_-$, this ergoregion extends to infinity near
to the rotation axis.

\section{Comparison with the Linearised Wald Analysis}

   It is instructive to compare our results with those that Wald obtained 
\cite{Wald} by employing a linearised analysis starting with the Kerr solution.
Wald's analysis ignored the back reaction of the magnetic field on the
Kerr metric.  The back reaction becomes important at radii $\rho$ that
are comparable to or greater than $\rho_{\rm Melvin}$, defined in
(\ref{melvinradius}). 
As long as the horizon radius is much smaller than $\rho_{\rm Melvin}$, 
i.e.\ $m<<B^{-1}$, the metric at distances large compared with the
horizon radius, but still much smaller than $\rho_{\rm Melvin}$, is well
approximated by an asymptotically flat metric, as assumed in
Wald's discussion.  Therefore to make the comparison with our
results, which include 
without approximation the non-linear effects due to the back reaction, 
we may linearise our expressions for the magnetised
Kerr-Newman solution (with $p=0$).  Thus, we treat $q$ and $B$ as small,
and keep only terms up to linear order in small quantities. Stated
precisely, we rescale $q\rightarrow k q$ and $B\rightarrow k B$ in the exact
solution, expand up to linear order in $k$, and then set $k=1$.  In this
approximation the metric becomes precisely the uncharged Kerr metric, and
the gauge potential, after making the gauge transformation $A\rightarrow A +
  q/(2m)\, dt$ for convenience, becomes
\be
A^{\rm lin} = -\fft{q}{2m}\, K_\flat + \ft12 
                       B \, {\mathfrak m}_\flat\,,\label{Alin1}
\ee
where $K_\flat= g_{t\mu} dx^\mu$ and ${\mathfrak m}_\flat= g_{\phi\mu} dx^\mu$.
Using the expression (\ref{Qcharge}) for the physical charge $Q$ on the black
hole, which becomes, after linearisation, $Q=q+2amB$, and using the fact that
the angular momentum of the Kerr black hole is given by $j=am$, we see that
(\ref{Alin1}) becomes 
\be
A^{\rm lin} = \fft{(2 j B-Q)}{2m} \, K_\flat + 
             \ft12 B \, {\mathfrak m}_\flat\,.
\label{Alin2}
\ee
Note that $j$ may be evaluated by a Komar integral over a surface at a
radius much larger than the horizon radius, but still much smaller than the
Melvin radius $2/B$.  There is no obvious relation between $j$ and a  
(possibly
regularised) Komar integral taken over a surface whose radius is much
greater than the Melvin radius.

\subsection{The First and Second Laws, and injection energies}

Let us suppose that for some choice of timelike Killing vector field
$K^\mu$  the future-directed 
null generator $l^\mu $ of the horizon is given by 
\ben
l^\mu = K^\mu + \Omega _H m^\mu \,,
\een
where $\Omega_H$ is the angular velocity
of the horizon.  The future-directed  mechanical 4-momentum $p_\mu$ of an 
infalling particle of mass $m$ and charge $q$   is given by   
\ben
p^\mu = m \frac{dx^\mu}{d \tau} \,.  
\een
It follows that
\ben
l^\mu p_\mu <0\,. 
\een 
Now the canonical 4-momentum  $\pi_\mu$ is given by   
\ben
\pi_\mu = p_\mu + q A_\mu 
\een
and hence
\ben
l^\mu \pi _\mu +q \Phi_H < 0
\een
where $\Phi_H =-l^\mu A_\mu $ is the electrostatic potential of the horizon,
and where a gauge  must be chosen which is regular on the horizon.
It is then known that this quantity is constant on the horizon.   
Now  $E_p=- K^\mu \pi_\mu $ is the conserved energy (with respect  to $K^\mu$),
and $J_p = m^\mu \pi_\mu$ is the conserved angular momentum of the 
infalling particle.
Thus  
\ben
E_p -\Omega_ H  J_p - \Phi_H\, q  >0 \,.   
\een

If we identify $E_p$ with $dE$, the gain  in energy of the horizon;
$J_p$ with $dJ$, the change in angular momentum  of the horizon;
and $q$ with $dQ$, the change in electric charge  of the horizon,
we shall have
\ben
dE -\Omega_ H  dJ  - \Phi_H  dQ  >0 \,. \label{laws}    
\een

In the asymptotically flat  case considered by Wald, the left-hand side of 
(\ref{laws}) equals $TdS = \frac{1}{8 \pi} \kappa   dA $,  
and we might expect this still to be true in the non-asymptotically flat
case for suitable definitions of $E$ and $J$.  

Now for a particle  falling along the axis, on which $m^\mu =0$,
we have $J_p=0$, and so the injection energy is determined
by  $\Phi_H dQ$, and in Walds's case $\Phi_H$ is determined
by the difference of  $-A_\mu K^\mu $ between the horizon and infinity.   
By (\ref{Alin2}) $\Phi_H\, dQ$ is given by
\ben
dQ (\frac{Q}{2m}- \frac{Bj}{m} ) \,.
\een
In our case, the analogue of Wald's  injection energy
would be proportional to  the difference of $-A_\mu K^\mu$ evaluated on the axis
between the horizon and infinity for an appropriate choice
of Killing vector $K^\mu$.

   Taking $\Omega=\Omega_H$, the angular velocity of the horizon given by
(\ref{OmegaH}), we find that on the horizon,
\be
(-A_\mu K^\mu_{\Omega_H})|_H = \fft{a B(1 - 2 m^2 B^2 -\ft14(11+\tvep^2)
  m^4 B^4)}{(1+m^2 B^2)}\,.
\ee
At large distances, we find
\be
-A_\mu K^\mu_{\Omega_H} = \ft12 a m B^3 (1+\cos^2\theta)\, r + {\cal O}(r^0)\,.
\label{large}\ee
 
Because $A _\mu K^\mu_{\Omega_H}$ diverges at infinity 
(in a direction-dependent fashion), we cannot apply
Wald's injection energy argument. In his case, which can be obtained from
our results by dropping terms beyond the linear order in $B$, 
the quantity $A _\mu K^\mu_{\Omega_H}$ in fact
tends to zero at infinity and so the difference between its value
on the horizon (where it is constant) and at infinity is well
defined and finite. Since Wald's calculation and ours are essentially
different, it is perhaps not surprising that we obtain a different result.

\section{Conclusion}

In this paper we have resolved some longstanding puzzles
concerning the behaviour of a magnetized Kerr-Newman black hole
immersed in an external
magnetic field $B$ which have previously been obscured
by the algebraic complexity of the exact metrics, for which we 
give a complete and self-contained derivation. 

Specifically we have identified 
the  criterion $Q= jB(1+\ft14 j^2 B^4)$ 
which must be satisfied if the metric is to be asymptotic to the Melvin metric
and there is to be no ergoregion 
associated with the  Killing vector field 
$K_\Omega = \frac{\partial}{\partial t} + 
\Omega \frac{\partial}{\partial \phi}$  
in the neighbourhood
of infinity for some choice of angular velocity
$\Omega$.\footnote{Note that $Q$ is the conserved electric charge calculated
in the exact geometry of the magnetised black hole.  However, $j=am$ is
the angular momentum of the original Kerr-Newman seed solution.  It
is not clear at present how one might calculate the true conserved
angular momentum of the magnetised black hole.  An equivalent statement of the
criterion is $q=-amB$, where $q$ is the charge parameter of the seed 
solution.} 
If this charge criterion is satisfied, the electromagnetic field has both
electric and magnetic components that are asymptotically constant along the 
axis.  The solution is asymptotic to a duality rotation of the Melvin
magnetic universe.  There are then 
two natural rigidly-rotating  frames of reference that are of particular
interest, 
associated with the  Killing vector field 
$K_\Omega$.    One, which has $\Omega=\Omega_s$ (see (\ref{Omegasol})),
may be thought of as non-rotating near infinity. In this case
there is, in general, an ergoregion confined to a neighbourhood
of the horizon.  The other frame, for which $\Omega=\Omega_H$ 
(see (\ref{OmegaH})), 
may be thought of as co-rotating with the horizon. In this case,
again provided our criterion $q=-amB$ is satisfied, numerical studies 
indicate there is no 
ergoregion at all if $B$ is sufficiently large ($B>B_+$, defined in
section 3.2): the associated Killing vector field $K_\Omega$ 
is everywhere timelike  outside the horizon.  For $B$ in the range $B_-<B<B_+$
(where $B_-$ is also defined in section 3.2), there is a toroidal ergoregion
outside and disjoint from the horizon.  If $B\le B_-$ this ergoregion extends
out to infinity in a tubular region near to the rotation axis.

This somewhat non-intuitive behaviour,
which is not encountered in the asymptotically 
flat case which is recovered in the absence of a magnetic field,
may be attributed to the highly curved geometry near infinity
which results from the full non-linear back-reaction, and 
is analogous to a similar phenomenon encountered
in the case of AdS black holes.   

Our criterion charge differs from the condition $Q=2jB$ obtained 
by Wald on energetic grounds. This may also be
attributed  to the difference between the fields
when back-reaction is taken into account. 
One may show how Wald's results can be recovered  by
measuring energies not with respect to infinity, but in an intermediate
region whose distance is large compared with the horizon
radius but small compared with the Melvin radius, $\frac{2}{B}$, 
at which substantial back-reaction effects set in. 
                
The ultimate aim of the work reported here is to 
bring  to bear the machinery of black hole thermodynamics 
at the fully non-linear level 
to the  physically important problem of the energetics of black holes 
in magnetic fields. Our results leave many questions
unanswered, but without a good understanding  of
the appropriate reference systems  to use, progress 
is  blocked. The results of this paper illustrate
the subtlety of the problem.

\section*{Acknowledgements}

We are grateful to Ji\v r\`i Bi\v c\'ak and Donald Lynden-Bell 
for helpful conversations. 
The research of C.N.P. is supported in part by
DOE grant DE-FG03-95ER40917.

\appendix

\section{Magnetising Transformation}

  The procedure that we shall use for generating the magnetised black hole
solutions makes use of the global $SU(2,1)$ symmetry group that emerges after
performing a Kaluza-Klein reduction of the four-dimensional
Einstein-Maxwell action and dualising the vector fields to scalars 
in three dimensions.  Similar techniques are described in, for example,
\cite{exact,bregibmai,cjlp3}.

  We begin with the four-dimensional Einstein-Maxwell theory described by
the Lagrangian
\be
{\cal L}_4 = \hat R-  \hat F^2\,.
\ee
(All four-dimensional quantities are hatted.)  We consider a solution with
\bea
d\hat s_4^2 &=& e^{2\varphi}\, ds_3^2 + e^{-2\varphi} (dz +2\cA)^2\,,\nn\\
\hat A &=& A + \chi\, (dz+2 \cA)\,,\label{kkred}
\eea
where all quantities on the right-hand sides are independent of $z$.  The
reduced three-dimensional Lagrangian is given by
\be
{\cal L}_3 = R - 2 (\del\varphi)^2- 2 e^{2\varphi} (\del\chi)^2 
             - e^{-4\varphi} {\cal F}^2 
   -e^{-2\varphi} F^2  \,,
\ee
where
\be
\cF=d\cA\,,\qquad F=dA + 2 \chi\, d\cA \,.
\ee

Adding Lagrange multipliers $4d\psi\wedge (F-2\chi \cF) + 4d\sigma\wedge\cF$
and eliminating $F$ and $\cF$, we obtain the dualised Lagrangian
\be
{\cal L}_3 =R- 2(\del\varphi)^2 - 2 e^{2\varphi} (\del\chi)^2 -
  2 e^{2\varphi} (\del\psi)^2 - 2 e^{4\varphi} (d\sigma-2\chi d\psi)^2\,.
\label{su21lag}
\ee
The two formulations are related by
\be
e^{-2\varphi} {*F}= d\psi\,,
 \qquad e^{-4\varphi} {*\cF} =d\sigma-2 \chi d\psi\,.
\ee
This implies that
\be
\hat F= - e^{2\varphi} {*d\psi} + d\chi\wedge (dz+2\cA)\,.\label{hatF}
\ee

The sigma model metric
\be
d\Sigma^2 = d\varphi^2 + e^\varphi (d\chi^2 +d\psi^2) + 
   e^{2\varphi} (d\sigma-\chi d\psi)^2
\ee
is the Fubini-Study metric on the non-compact 
$\widetilde {\CP^2} =SU(2,1)/U(2)$, with $R_{ij}=-\ft32 g_{ij}$. It has
the K\"ahler form 
\be
J=e^\varphi\, [d\varphi\wedge (d\sigma-\chi d\psi) +d\psi\wedge d\chi]
= d[e^\varphi (d\sigma-\chi d\psi)]\,.
\ee

Defining the $3\times 3$ matrices $E_a{}^b$ to have zeroes everywhere except 
for a 1 at row $a$, column $b$, we can parameterise a coset representative
as
\be
\cV= e^{\varphi H} e^{-\im \sigma E_0{}^2} 
  e^{\sqrt2\, \chi (E_0{}^1 + E_1{}^2)}
       e^{-\im\sqrt2\, \psi(E_0{}^1 - E_1{}^2)}\,,
\ee
where $H=E_0{}^0- E_2{}^2$. It can be verified that $\cV$ is in $SU(2,1)$, 
with
\be
\cV^\dagger \eta \cV = \eta\,,\qquad \eta=\begin{pmatrix} 0& 0& -1\\
                                                   0 & 1 & 0\\
                                                   -1 & 0 & 0
                                           \end{pmatrix}\,,
\ee
with $\eta$ being the invariant metric of $SU(2,1)$.  Defining
\be
{\cal M} = \cV^\dagger \cV\,
\ee
the Lagrangian (\ref{su21lag}) can be written as
\be
{\cal L}_3 = R -{\rm tr}({\cal M}^{-1}\del{\cal M})^2\,.
\ee
This makes manifest that ${\cal L}_3$ is invariant under $SU(2,1)$, with
\be
{\cal M} \longrightarrow {\cal M}'=U^\dagger {\cal M} U\,,\label{Utrans}
\ee
where $U$ is any constant $SU(2,1)$ matrix, obeying $U^\dagger \eta U=\eta$.

   The specific $SU(2,1)$ transformation that generates magnetised solutions
from non-magnetised ones is given by taking
\be
U= \begin{pmatrix} 1&0&0\\
                   \fft{B}{\sqrt2}  & 1 & 0\\
                  \fft{B^2}{4} & \fft{B}{\sqrt2} & 1
    \end{pmatrix}\,.\label{Umatrix}
\ee

More generally, we can generate solutions with an external electric field
$E$ and magnetic field $B$ using
\be
U= \begin{pmatrix} 1&0&0\\
                   \fft{(B+ \im E)}{\sqrt2}  & 1 & 0\\
                  \fft{(B^2+E^2)}{4} & \fft{(B-\im E)}{\sqrt2} & 1
    \end{pmatrix}\,.\label{UmatrixEB}
\ee

  Electric/Magnetic duality in four dimensions corresponds to a $U(1)$ 
rotation 
\bea
\hat F\longrightarrow \hat F' &=& \hat F \cos\alpha + 
   {\hat * \hat F}\sin\alpha\,,
\nn\\
&=& -e^{2\varphi} {*d\psi'} + d\chi'\wedge(dz+2\cA)\,,\label{hattF}
\eea
where
\be
\begin{pmatrix} \chi' \\\psi' \end{pmatrix} = 
\begin{pmatrix} \cos\alpha & \sin\alpha\\ -\sin\alpha &\cos\alpha \end{pmatrix}
\begin{pmatrix} \chi \\\psi \end{pmatrix}\,.
\ee
At the same time, maintaining the invariance of $d\sigma-\chi d\psi$ 
requires transforming $\sigma$ to
\be
\sigma' = \sigma + \ft12 \sin2\alpha\, (\psi^2-\chi^2) - 2\sin^2\alpha\, 
  \chi\psi\,.
\ee
This duality transformation is implemented on $\cM$ by (\ref{Utrans}) with
the $U(1)\in SU(2,1)$ matrix
\be
U =\begin{pmatrix} e^{-\ft{\im}{3}\alpha} & 0 & 0\\
                0 & e^{\ft{2\im}{3}\alpha} & 0\\
                0 & 0 & e^{-\ft{\im}{3} \alpha} \end{pmatrix}\,.
\ee

  The complex Ernst potential $\Phi$ is defined by $d\Phi=i_K({{\hat *}\hat F}
 +\im\, \hat F)$, where $K=\del/\del z$ and $i_K \omega= 
 K^\mu \omega_{\mu\nu} dx^\nu$ for any 2-form $\omega$. 
From (\ref{hatF}) and (\ref{hattF}),
we see that we can take
\be
\Phi= \psi+\im\, \chi\,.
\ee

\section{Magnetised Kerr-Newman Metric}

  We begin with the Kerr-Newman solution describing a rotating black hole 
carrying an electric charge $q$ and a magnetic charge $p$.  It is given by
\bea
d\hat s_4^2 &=& - fdt^2 + 
   R^2 \Big( \fft{dr^2}{\Delta} + d\theta^2\Big)
+ \fft{\Sigma\, \sin^2\theta}{R^2}\,  (d\phi-\bar\omega dt)^2\,,\nn\\
A &=& \bar\Phi_0\, dt + \bar\Phi_3\, (d\phi-\bar\omega dt)\,,
\eea
where
\bea
R^2&=&r^2+a^2\cos^2\theta\,,\qquad 
        \Delta= (r^2+a^2) - 2m r + q^2+p^2\,,\nn\\
\bar\omega&=& \fft{a(2mr-q^2-p^2)}{\Sigma}\,,\qquad
  f= \fft{R^2\Delta}{\Sigma}\,,\qquad
\Sigma=(r^2+a^2)^2 - a^2\Delta\sin^2\theta\,,\label{knfns}
\eea
and
\bea
\bar \Phi_0 &=& - \fft{q r (r^2+a^2)}{\Sigma} + 
                 \fft{ ap \Delta\cos\theta}{\Sigma}
\,,\nn\\
\bar\Phi_3 &=& \fft{a q r \sin^2\theta}{R^2} - 
                 \fft{p (r^2+a^2)\cos\theta}{R^2}\,.
\eea

   After applying the procedure described previously with the transformation
(\ref{Umatrix}), we arrive at the magnetised Kerr-Newman solution
\bea
d\hat s_4^2 &=&H\,\Big[- fdt^2 +
   R^2 \Big( \fft{dr^2}{\Delta} + d\theta^2\Big)\Big]
+ \fft{\Sigma\, \sin^2\theta}{H\, R^2}\,  (d\phi-\omega dt)^2\,,\nn\\
A &=& \Phi_0\, dt + \Phi_3\, (d\phi-\omega dt)\,,\label{magkn}
\eea
where
\be
H = 1 +\fft{H_\1 B + H_\2 B^2+ H_\3 B^3 + 
         H_\4 B^4}{R^2}\,,
\ee
with
\bea
H_\1 &=& 2 a q r \sin^2\theta - 
                  2 p(r^2+a^2) \cos\theta\,,\nn\\
H_\2 &=&\ft12[(r^2+a^2)^2 - a^2 \Delta\, \sin^2\theta]\, \sin^2\theta +
      \ft32 \tilde q^2(a^2+ r^2\cos^2\theta)\,,\nn\\
H_\3&=& - p a^2  \Delta  \sin^2\theta\cos\theta - \fft{q a\Delta}{2r} 
         [r^2(3-\cos^2\theta)\cos^2\theta + a^2(1+\cos^2\theta)] 
+\fft{a q(r^2+a^2)^2 (1+\cos^2\theta)}{2r} \nn\\
&&-
            \ft12 p(r^4-a^4)\sin^2\theta\cos\theta
  +\fft{q \tilde q^2 a [(2r^2+a^2) \cos^2\theta + a^2]}{2r} -
  \ft12 p \tilde q^2 (r^2+a^2)\cos^3\theta\,,\nn\\
H_\4 &=& \ft1{16} (r^2+a^2)^2 R^2 \sin^4\theta +
                \ft14 m a^2 r (r^2+a^2)\sin^6\theta +
   \ft14 m a^2\tilde q^2 r (\cos^2\theta -5)\sin^2\theta\cos^2\theta\nn\\
&& 
  +\ft14 m^2 a^2[ r^2 (\cos^2\theta-3)^2 \cos^2\theta + a^2(1+\cos^2\theta)^2]
  \\
&& +\ft18\tilde q^2 (r^2+a^2)(r^2 +a^2 + 
         a^2\sin^2\theta)\sin^2\theta\cos^2\theta +
\ft1{16} \tilde q^4 [r^2\cos^2\theta + a^2 (1+\sin^2\theta)^2]\cos^2\theta\,,
\nn
\eea
and we have defined 
\be
\tilde q^2 \equiv q^2+p^2\,.
\ee
The quantity $\omega$ is given by
\be
\omega=\fft{(2m r-\tilde q^2)a + \omega_\1 B +
\omega_\2 B^2 + \omega_\3 B^3 + \omega_\4 B^4}{\Sigma}\,,\label{omegakn}
\ee
where 
\bea
\omega_\1 &=& -2 q r(r^2+a^2) + 2 a p \Delta\cos\theta\,,\nn\\
\omega_\2 &=& -\ft32 a \tilde q^2 (r^2+a^2 + \Delta\cos^2\theta)\,,\nn\\
\omega_\3 &=& 4 q m^2 a^2 r + \ft12 a p \tilde q^4 \cos^3\theta +
   \ft12 q r (r^2+a^2)[r^2-a^2 + (r^2+3a^2)\cos^2\theta] \nn\\
&&+
  \ft12 a p (r^2+a^2) [3r^2+a^2 -(r^2-a^2)\cos^2\theta]\cos\theta
  +\ft12 q \tilde q^2 r [(r^2+3a^2)\cos^2\theta - 2a^2] \nn\\
&&+
  \ft12 a p \tilde q^2[ 3r^2+a^2 + 2 a^2 \cos^2\theta]\cos\theta
-a m \tilde q^2 (2 a q + p r \cos^3\theta)\nn\\
&& 
  +q m[r^4-a^4 + r^2(r^2+3 a^2)\sin^2\theta] 
  -a p m r [2 R^2 + (r^2+a^2)\sin^2\theta]\, \cos\theta \,,\nn\\
\omega_\4 &=& \ft12 a^3 m^3 r (3+\cos^4\theta) 
       -\ft1{16} a\tilde q^6 \cos^4\theta 
  -\ft18 a \tilde q^4 [r^2(2+\sin^2\theta)\cos^2\theta + a^2(1+\cos^4\theta)]
\nn\\
&&
+\ft1{16} a\tilde q^2 (r^2+a^2)[r^2(1-6\cos^2\theta + 3\cos^4\theta) -
           a^2(1+\cos^4\theta)] -\ft14 a^3 m^2 \tilde q^2 (3+\cos^4\theta) 
\nn\\
&&
+\ft14 a m^2 [r^4(3-6\cos^2\theta +\cos^4\theta) + 
             2 a^2 r^2 (3\sin^2\theta-2\cos^4\theta)- a^4 (1+\cos^4\theta)]
\nn\\
&&
+ \ft18 a m \tilde q^4 r \cos^4\theta 
+\ft14 a m \tilde q^2 r[2r^2(3-\cos^2\theta)\cos^2\theta 
                         -a^2(1-3\cos^2\theta-2\cos^4\theta)] \nn\\
&&
+
\ft18 a m r (r^2+a^2)[r^2(3+6\cos^2\theta -\cos^4\theta) 
                     -a^2(1-6\cos^2\theta-3\cos^4\theta)]\,.\label{omegakn2}
\eea

  Since
\bea
\hat F= -e^{2\varphi} {*d\psi} + d\chi\wedge (d\phi-\omega dt)\,,\qquad
{{\hat *} \hat F}= e^{2\varphi}{*d\chi} + d\psi\wedge(d\phi-\omega dt)\,,
\eea
the conserved electric and magnetic charges, obtained by integrating
the 2-forms ${{\hat *} \hat F} = {{\hat *} \hat F}_{23} d\theta\wedge d\phi
+\cdots$
and $\hat F= \hat F_{23} d\theta\wedge d\phi+\cdots$ over a 2-sphere, can
be determined from the knowledge of $\psi$ and of $\chi=\Phi_3$ respectively.
In fact, comparing with the papers of Ernst et al., we find that the 
complex Ernst potential $\Phi$ is given by
\be
\Phi = \psi + \im\, \chi\,.
\ee
In particular, we will have, in vierbein components,
\bea
{{\hat *}\hat F}_{23} + \im \hat F_{23} = E_r + \im H_r &=&
  \fft{f^{1/2}}{R\Delta^{1/2}\,\sin\theta}\, \fft{\del}{\del\theta}
  (\psi +\im\, \chi)\,,\nn\\
-{{\hat *}\hat F}_{13} - \im \hat F_{13} = E_\theta + \im H_\theta &=&-
  \fft{f^{1/2}}{R\,\sin\theta}\, \fft{\del}{\del r}
  (\psi +\im\, \chi)\,.\label{vierbeinEH}
\eea

We find that for the magnetised Kerr-Newman solution,
\be
\psi= \fft{\psi_\0 + \psi_\1 B + \psi_\2 B^2}{R^2 H}\,,
\ee
where
\bea
\psi_\0 &=&  q (r^2+a^2)\cos\theta +  a p r \sin^2\theta\,,\nn\\
\psi_\1 &=&  a m [3 r^2 + a^2 -(r^2-a^2)\cos^2\theta]\cos\theta -
       a \tilde q^2 r \sin^2\theta\cos\theta\,,\nn\\
\psi_\2 &=&-\ft14 q (r^2+a^2)^2 \sin^2\theta\cos\theta - 
   \ft14 a p r (r^2+a^2)\sin^4\theta
 +  a^2 q m r \sin^2\theta\cos\theta \nn\\
&&
-
   \ft12 a p m[ r^2 (3-\cos^2\theta)\cos^2\theta + a^2(1+\cos^2\theta] 
  -\ft14 q \tilde q^2 [(r^2-a^2)\cos^2\theta + 2 a^2]\cos\theta
 \nn\\
&&
 + \ft14 a p \tilde q^2 r \sin^2\theta\cos^2\theta\,.
\eea
The potential $\Phi_3= \chi$ is given by
\be
\Phi_3=\chi = \fft{\chi_\0 + \chi_\1 B + \chi_\2 B^2 + \chi_\3 B^3}{R^2 H}\,,
\label{Phi3kn}
\ee
where
\bea
\chi_\0 &=& a q r \sin^2\theta - p (r^2+a^2)\cos\theta\,,\nn\\
\chi_\1 &=& \ft12[\Sigma \sin^2\theta + 
              3 \tilde q^2 (a^2+ r^2\cos^2\theta)]\,,\nn\\
\chi_\2 &=& \ft34 a q r (r^2+a^2)\sin^4\theta 
   -\ft34 p (r^2+a^2)^2 \sin^2\theta\cos\theta
+3 a^2 p m r \sin^2\theta\cos\theta 
\nn\\
&&
  +\ft32 a q m[r^2 (3-\cos^2\theta)\cos^2\theta + a^2(1+\cos^2\theta)] -
\ft34 a q \tilde q^2 r \sin^2\theta\cos^2\theta \nn\\
&&
-
\ft34 p \tilde q^2 [(r^2-a^2)\cos^2\theta + 2 a^2]\cos\theta\,,\nn\\
\chi_\3&=& \ft18 R^2(r^2+a^2)^2 \sin^4\theta  
  + \ft12 a^2 m r (r^2+a^2)\sin^6\theta - 
 \ft12 a^2 \tilde q^2 m r (5-\cos^2\theta)\sin^2\theta\cos^2\theta\nn\\
&&
 + \ft12 a^2 m^2[ r^2 (3-\cos^2\theta)^2 \cos^2\theta + 
           a^2(1+\cos^2\theta)^2] \nn\\
&&
+
  \ft14 \tilde q^2 (r^2+a^2)[r^2 + a^2 + a^2\sin^2\theta]
                  \sin^2\theta\cos^2\theta\nn\\
&& 
   +\ft18 \tilde q^4 [r^2\cos^2\theta + a^2(2-\cos^2\theta)^2]\,\cos^2\theta\,.
\label{Phi3kn2}
\eea

The potential $\Phi_0$ is given by
\be
\Phi_0 = \fft{\Phi_0^\0 + \Phi_0^\1\, B + \Phi_0^\2\, B^2 +\Phi_0^\3\, B^3}{
                 4\Sigma}\,,\label{Phi0kn}
\ee
where
\bea
\Phi_0^\0 &=& 4[-q r(r^2+a^2) + a p \Delta \cos\theta]\,,\nn\\
\Phi_0^\1 &=& -6 a \tilde q^2(r^2+a^2 + \Delta \cos^2\theta)\,,\nn\\
\Phi_0^\2 &=& -3 q[ (r+2m)a^4 - (r^2+4mr + \Delta \cos^2\theta) r^3
   + a^2(2\tilde q^2(r+2m) -6m r^2 -8m^2 r \nn\\
&&
\qquad - 3\Delta r \cos^2\theta)] +
      3 p \Delta [3 a r^2 + a^3 + a(a^2+\tilde q^2 -r^2)\cos^2\theta]\cos\theta
\,,\label{Phi0kn2}\\
\Phi_0^\3 &=& -\ft12 a\Big\{ 4 a^4 m^2 + a^4 \tilde q^2 + 12 a^2 m^2 \tilde q^2 +
     2 a^2 \tilde q^4 + 2 a^4 m r
 - 24 a^2 m^3 r + 4 a^2 m \tilde q^2 r \nn\\
&&
- 24 a^2 m^2 r^2 - 4 a^2 m r^3 - 12 m^2 r^4 - \tilde q^2 r^4 - 6 m r^5
 -6 r\Delta [2m(r^2+a^2) -\tilde q^2 r]\cos^2\theta \nn\\
&&+
 \Delta (\tilde q^4 - 3\tilde q^2 r^2 + 2 m r^3 + 
a^2(4m^2 +\tilde q^2 - 6 m r)]
   \cos^4\theta\Big\}\,.\nn
\eea

  The vector potential generating the electromagnetic
field is given by (\ref{magkn}), using (\ref{omegakn}), (\ref{omegakn2}), 
(\ref{Phi3kn}), (\ref{Phi3kn2}), (\ref{Phi0kn}) and (\ref{Phi0kn2}).  In 
principle, this could be used to calculate the orbits of charged particles
in the magnetised Kerr-Newman solution.

\section{Generating Taub Cosmological Metric}

   A ``group commutator'' $U(E,0)^{-1} U(0,B)^{-1} U(E,0) U(0,B)$, where
$U(E,B)$ is given by (\ref{UmatrixEB}), takes the form
\be
U = \begin{pmatrix} 1&0&0\\  0&1&0\\ \im c &0&1 \end{pmatrix}\,,
\label{Uc}
\ee
where $c=E B$.  Starting from the flat space metric
\be
d\bar s_4^2 = -dt^2 + dr^2 + r^2 d\theta^2 + r^2\sin^2\theta d\phi^2
\ee
and acting with the $SU(2,1)$ transformation (\ref{Uc}), we obtain the
Ricci-flat metric 
\be
ds_4^2 = (1+ c^2 \rho^4)(-dt^2 + dz^2 + d\rho^2) + \fft{\rho^2}{1+c^2 \rho^4}\, 
  (d\phi - 4 c z dt)^2\,,\label{tmetric}
\ee
where we have defined $\rho=r\sin\theta$, $z=r\cos\theta$.  This is an
analytic continuation of a Bianchi II cosmological metric originally 
obtained by Taub \cite{Taub}.

   The metric (\ref{tmetric}) has
\be
-g_{00} = (1+c^2 \rho^4)^{-1}\, [(1+c^2 \rho^4)^2 - 16 c^2 z^2 \rho^2]\,.
\ee
Thus there is an ergoregion when
\be
|4c z \rho| > 1+c^2 \rho^4\,.
\ee

\end{document}